\documentclass[9pt,twocolumn,twoside]{optica}
\setboolean{shortarticle}{false}
\pdfoutput=1
\usepackage[mediumspace,mediumqspace,Grey,squaren]{SIunits}
\usepackage{bm}
\usepackage[utf8]{inputenc}
\usepackage{float}

\newcommand{\bbt}{\boldsymbol{\Theta}}
\newcommand{\bbtz}{\boldsymbol{\Theta_z}}
\newcommand{\bbta}{\boldsymbol{\Theta_a}}
\newcommand{\btp}{\boldsymbol{\Theta_{\textrm{paw}}}}
\newcommand{\br}{\boldsymbol{\rho}}
\newcommand{\brp}{\boldsymbol{\rho'}}
\newcommand{\bro}{\boldsymbol{\rho_0}}
\newcommand{\brc}{\boldsymbol{\rho_c}}
\newcommand{\brd}{\boldsymbol{\rho_d}}
\newcommand{\broc}{\boldsymbol{\rho_{0}}}
\newcommand{\bF}{\boldsymbol{F}}
\newcommand{\bFz}{\boldsymbol{F_z}}
\newcommand{\bnabla}{\boldsymbol{\nabla}}
\newcommand{\bs}{\boldsymbol{\hat{s}}}
\newcommand{\dV}{\textbf{V}_{\textrm{dm}}}
\newcommand{\bMp}{\textbf{M}^{\boldsymbol{+}}}

\title{Conjugate adaptive optics in widefield microscopy with an extended-source wavefront sensor}

\author[1]{Jiang Li}
\author[2]{Devin R. Beaulieu}
\author[2]{Hari Paudel}
\author[1]{Roman Barankov}
\author[2]{Thomas G. Bifano}
\author[1,2,*]{Jerome Mertz}

\affil[1]{Department of Biomedical Engineering, Boston University, 44 Cummington Mall, Boston, Massachusetts 02215, USA}
\affil[2]{Photonics Center, Boston University, 8 Saint Mary's St., Boston, Massachusetts 02215, USA}

\affil[*]{Corresponding author: jmertz@bu.edu}

\dates{Compiled \today}

\ociscodes{(110.1080) Active or adaptive optics; (100.5070) Phase retieval; (110.0180) Microscopy.}

\doi{\url{...}}

\begin{abstract}

Adaptive optics is a strategy to compensate for sample-induced aberrations in microscopy applications. Generally, it requires the presence of "guide stars" in the sample to serve as localized reference targets. We describe an implementation of conjugate adaptive optics that is amenable to widefield (i.e. non-scanning) microscopy, and can provide aberration corrections over potentially large fields of view without the use of guide stars. A unique feature of our implementation is that it is based on wavefront sensing with a single-shot partitioned-aperture sensor that provides large dynamic range compatible with extended samples. Combined information provided by this sensor and the imaging camera enable robust image de-blurring based on a rapid estimation of sample and aberrations obtained by closed-loop feedback. We present the theoretical principle of our technique and proof of concept experimental demonstrations.        

\bigskip
\bigskip

\end{abstract}

\setboolean{displaycopyright}{false}

\begin{document}

\maketitle
\thispagestyle{fancy}
\ifthenelse{\boolean{shortarticle}}{\abscontent}{}

\section{Introduction}

Sample-induced aberrations generally lead to reduced image quality in optical microscopy. A standard approach to counter such aberrations is to use adaptive optics (AO), which was first developed in astronomy \cite{tyson} but is now gaining traction in microscopy \cite{kubby_book, booth_review}. The basic idea of AO is to insert an active optical correction element, typically a deformable mirror, in the optical path of the microscope to compensate for the aberrations produced by sample. The most common placement of this correction element, by far, is in a pupil plane of the microscope optics, called pupil AO. However, as first recognized by the astronomy community \cite{beckers}, a placement of the correction element in a plane conjugate to a primary sample aberration plane can lead to a significant field of view (FOV) advantage when these aberrations are spatially varying. More recently this advantage of conjugate AO has been recognized by the microscopy community both in simulation studies \cite{sedat, booth, cui_wf} and in experiment \cite{retina, conjugate}. In this paper we describe a novel implementation of conjugate AO, bearing in mind that our results can be equally applied to pupil AO.       

In practice, two strategies have been employed to determine the actual wavefront to be applied to the correction element. The first involves optimizing a particular metric of the image itself \cite{girkin,debarre,ji,mosk,cui}. For example, in a linear microscopy application (the only application we consider here) a commonly used metric is image contrast. Different wavefronts are applied to the correction element and image contrast is maximized by an iterative procedure based on trial and error. An advantage of image-based AO is that it is simple to implement since it requires no additional hardware besides the correction element itself. A disadvantage is that the iteration procedure can be slow, making it difficult to implement in real time. A more serious disadvantage comes from difficulties in convergence. For example, there are many ways to increase image contrast that do not improve image quality at all, simply by manipulating light distributions. In practice, image-based AO works well when optimizing the contrast of well-defined isolated reference points (called "guide stars", a term borrowed from astronomy parlance \cite{tyson}), but does not work well when optimizing the contrast of a distributed object scene.

The second strategy to determine what wavefront to apply to the correction element makes use of a wavefront sensor \cite{miller,wilson,denk,kubby,betzig}. This has the advantage that it does away with the iterative guesswork associated with image-based AO, readily enabling real-time operation. But it has the disadvantage that it requires additional hardware, namely a sensor capable of directly measuring optical wavefronts. The most commonly used sensor is the Shack-Hartmann (SH) wavefront sensor \cite{shack}, which has the benefit of being achromatic, meaning it can be used with quasi-broadband light (e.g. fluorescence). But a SH sensor exhibits both poor spatial resolution and limited dynamic range. The latter constraint means it has poor tolerance to angular diversity and can only be operated with quasi-collimated light. In effect, this too imposes the requirement of a well defined guide star in the sample.    

In this work we demonstrate an implementation of widefield microscopy with sensor-based AO that does not require the use of guide stars. Wavefront sensing is performed using illumination provided directly by the object itself, over the entire FOV of the wavefront correction. Since our implementation here involves conjugate AO (as opposed to pupil AO), the correction FOV is almost as large as the full FOV of our microscope. The development of our technique addressed two key challenges. The first challenge was the development of a wavefront sensor that exhibits large dynamic range capable of operating with relatively uncollimated light. For this we used a technique called partitioned aperture wavefront (PAW) sensing, previously developed in our lab for quantitative wavefront sensing both in transmission \cite{paw} and reflection \cite{roman} geometries. A PAW sensor is just as simple to operate as a SH sensor, and shares the same benefit of being achromatic. But unlike SH, a PAW sensor actually requires uncollimated light to function at all. The second challenge was to modify PAW sensing to enable it to work with an arbitrarily distributed extended source, namely the object itself. As we will see, this required supplementing PAW sensing with additional information provided by the science camera in our system (i.e. the imaging camera focused on the object). In this regard, our technique is similar to strategies involving joint estimation of object and aberrations \cite{fienup,allen,gureyev,
barbastasis,waller}, though it is faster and more direct. 

The layout of our paper is as follows. We first present the theoretical principles of our wavefront sensing strategy. This is followed by a description of our experimental setup and experimental results. As emphasized in the discussion, our results are confined here to partially coherent trans-illumination imaging with planar samples and aberrations. As such, they are intended to lay preliminary groundwork for future applications involving more general volumetric sample and aberrations.     

\section{Wavefront sensing with arbitrarily distributed extended sources}
Wavefront sensing requires the quantitative imaging of both the amplitude and phase of a wavefront. Since any standard camera provides amplitude imaging, the difficulty in wavefront sensing comes from phase imaging.  Several techniques are available for quantitative phase imaging \cite{popescu}, most of which are applicable only to monochromatic light. In the case of non-monochromatic or quasi-broadband light (e.g. fluorescence), optical phase is not well defined. Instead, what can be measured is changes in optical phase relative to a self-reference provided, for example, by spatial filtering \cite{slim, bernet} or shearing \cite{arnison}. The latter technique, in particular, provides access to the transverse gradient of the wavefront phase. This same quantity can be accessed alternatively by  measuring local tilt angles of the optical flux density, which is the strategy employed by SH sensors (and variations \cite{monneret}), and pyramidal wavefront sensors (e.g. \cite{iglesias}), of which PAW is an achromatic variation. Because of its advantages in spatial resolution and, more importantly, dynamic range, we will henceforth concentrate on PAW as our preferred wavefront sensing technology. In addition to these advantages, PAW is versatile (can be implemented with any standard microscope), robust (no moving parts), fast (single shot), light efficient (no requirement of pinholes), non-interferometric (speckle-free), and polarization independent. 

To date, we have demonstrated the effectiveness of PAW in cases where the illumination source, in addition to being extended, was uniform (as established by K\"ohler illumination) and symmetrically distributed about the optical axis. Our goal here is different. We wish to use the object itself as the illumination source for PAW sensing. This source is unknown in advance and, in general, arbitrarily distributed.   

\begin{figure}[htbp]
	\centering
	\makebox{\includegraphics[width=0.95\linewidth]{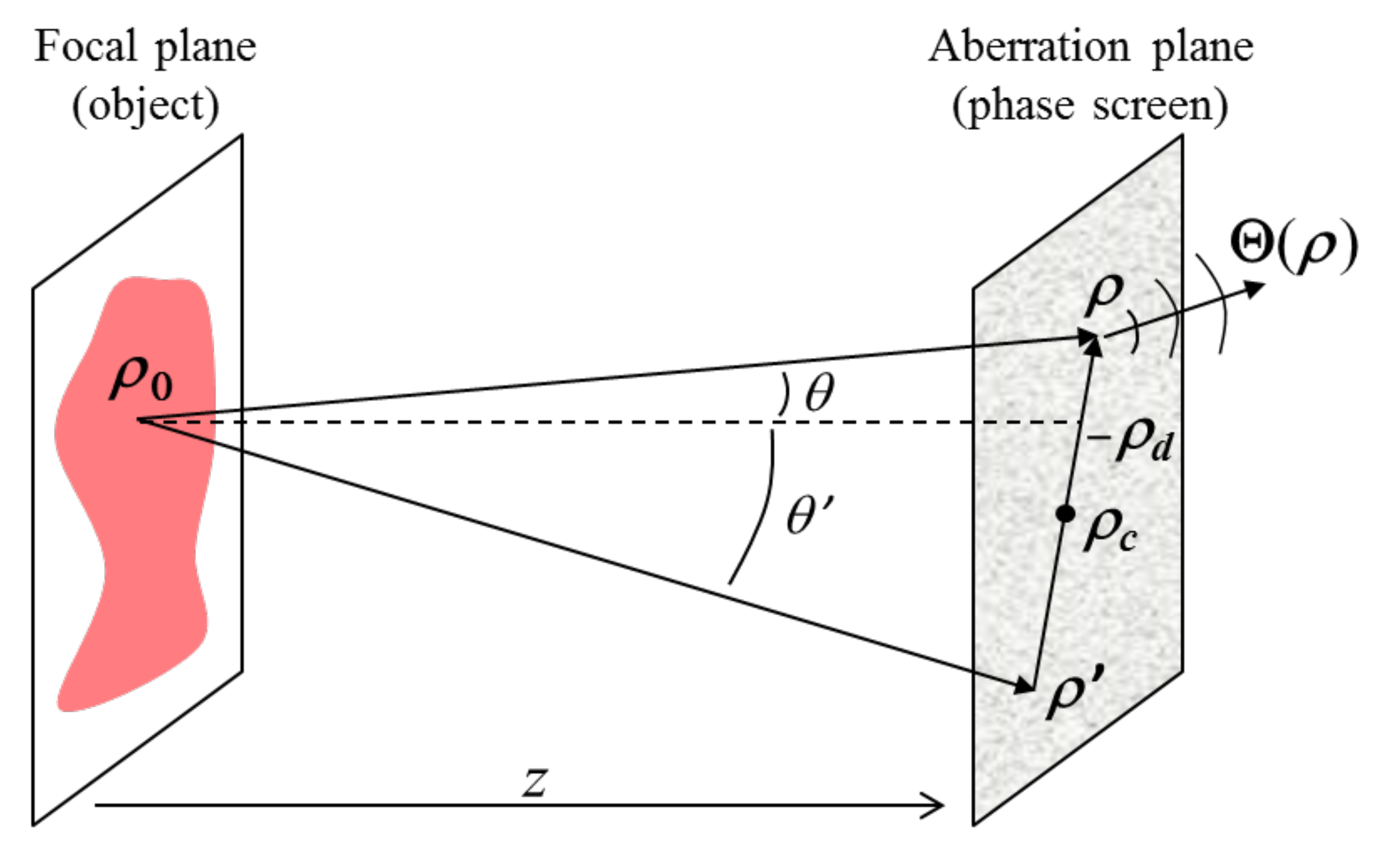}}
	\captionsetup{singlelinecheck=off}
	\caption{Geometry of focal and aberration planes.}
	\label{fig:figure1}
\end{figure}

The basic problem is depicted in Fig. \ref{fig:figure1} We consider the simplified case where a 2D object is located at the focal plane of our microscope, and a 2D phase screen is located at an out-of-focus plane a distance $z$ from the focal plane (extensions to more general cases will be discussed later). The object is taken to be incoherent, both spatially and temporally (e.g. a 2D distribution of fluorescent molecules). The phase screen is taken to be weakly scattering, imparting only paraxial tilt angle changes to the wavefront. This scenario has been investigated in detail, both theoretically and experimentally \cite{conjugate}. Our goal here is to measure the aberrations induced by the phase screen using only the illumination provided by the object, whose intensity $I_0(\bro)$ is arbitrarily distributed. To this end, we insert a PAW sensor in our system (not shown) that is focused onto the phase screen. The PAW sensor reveals both the local intensity $I(\br)$ emerging from the phase screen, and the average local tilt angle $\bbt(\br)$ of the flux density $\bF(\br)$. This last quantity is defined by \cite{ishimaru}
\begin{equation}
\bF(\br)=\int{L(\br, \bs)\bs\ d^2\bs},
\end{equation}
\noindent where $L(\br, \bs)$ is the light radiance at plane $z$ (or brightness or specific intensity), and $\bs$ is a unit direction vector, considered here with a net forward component, leading to $\bbt(\br)=\bF(\br)/I(\br)$.

As defined, $\bbt(\br)=(0,0)$ corresponds to at flux density at plane $z$ directed along the optical axis. In the absence of a phase screen this occurs when $I_0(\bro)$ is perfectly uniformly distributed. However, in general $I_0(\bro)$ is not uniformly distributed and hence $\bbt(\br)$ is not equal to zero, even in the absence of the phase screen. We thus have $\bbt(\br)=\bbtz(\br)+\bbta(\br)$, where $\bbtz(\br)$ is the flux density direction in the absence of the phase screen and $\bbta(\br)$ is the change in flux density direction induced by the phase screen. It is this change in flux density direction $\bbta(\br)$ that ultimately must be corrected by AO. But our PAW sensor supplies a measurement only of $\bbt(\br)$. Our problem therefore reduces to how to extract $\bbta(\br)$ from $\bbt(\br)$, or, said differently, how to first estimate $\bbtz(\br)$ (henceforth the subscript $z$ will denote parameters at plane $z$ in the absence of the phase screen).       

\begin{figure*}[t]
	\centering
	\makebox{\includegraphics[width=0.85\linewidth]{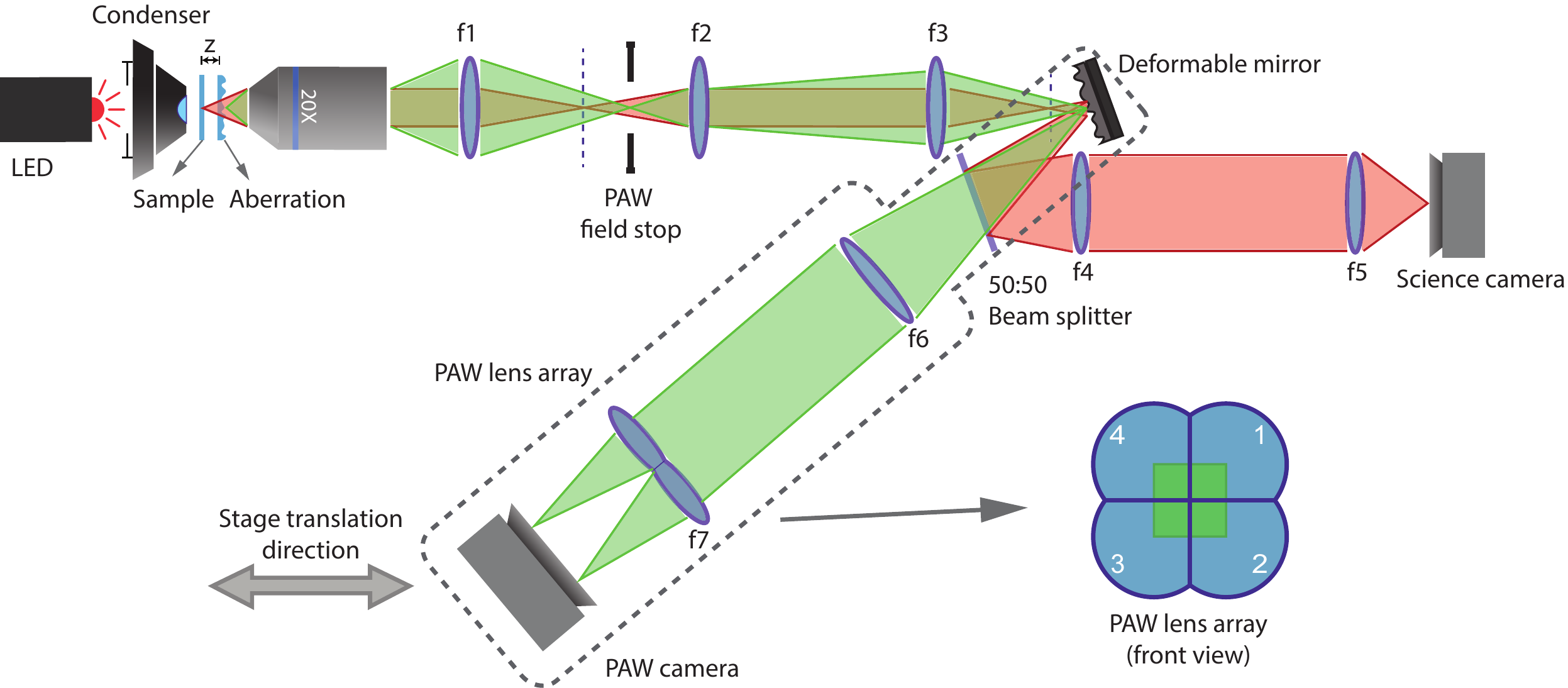}}
	\captionsetup{singlelinecheck=off}
	\caption{Experimental setup. A trans-illuminated sample followed by a phase screen is imaged onto a science camera (lenses f1-5) with magnification $4.6\times$ (imaging path in red; vertical dashed lines denote intermediate image planes). A deformable mirror (DM) is inserted into the optical path conjugate to the phase screen, and imaged with a PAW sensor comprising a main lens f6 and quatrefoil lens f7 (inset) in a 3f configuration (wavefront sensing path in green). A PAW field stop prevents overlap of the four oblique-detection images projected onto the PAW camera. The DM and PAW sensor are mounted on a translatable stage enabling adjustable conjugation. Lens focal lengths: f1 = 50mm, f2 = 100mm, f3 = 100mm, f4 = 300mm, f5 = 250mm, f6 = 200mm, f7 = 250mm. }
	\label{fig:setup}
\end{figure*}

To address this problem, we note that our PAW sensor is not operating in isolation. Another sensor is at work, namely the science camera focused on the object itself. As a result, the intensity distribution $I_0(\bro)$ at the focal plane is not completely unknown since it can be estimated directly from the science camera image. This image is only an estimate because of the blurring due to the phase screen. Nevertheless, it can be exploited to obtain an estimate of $\bbtz(\br)$. Our strategy to do this comes from the convergence of two basic principles in optics. The first principle is the well-known Van Cittert-Zernike (VCZ) theorem, which in a small angle approximation states that \cite{bw}
\begin{equation}
J_z(\brc,\brd)=\frac{1}{z^2}\int{I_0(\broc) e^{i \frac{2 \pi}{\bar{\lambda} z} (\brc-\broc) \cdot \brd }  \cos \theta \cos\theta' d^2 \broc}, 
\end{equation}
\noindent where $J_z(\brc,\brd)$ is the mutual intensity at plane $z$, $\bar{\lambda}$ is the average light wavelength, and we have made use of the centered coordinate system $\brc=(\br+\brp)/2$ and $\brd=(\br-\brp)$. The tilt angles $\theta$ and $\theta'$ are shown in Fig. \ref{fig:figure1}. In our small angle approximation we have $\cos\theta \cos\theta' =1/(1+|\brc-\broc |^2/z^2+\rho_d^2/4z^2)$. We note that the VCZ theorem is valid provided $I_0(\broc)$ is spatially incoherent, which is assumed here.

The second principle we will make use of is the fundamental link between coherence and radiometry provided by \cite{ishimaru}
\begin{equation}
J_z(\brc,\brd)=\int{L_z(\brc, \bs) e^{i \frac{2 \pi}{\bar{\lambda}} \brd \cdot \bs} d^2 \bs},
\end{equation}
\noindent which, from Eq. 1, leads directly to
\begin{equation}
\bnabla_{\brd} J_z(\brc,\brd) |_{\brd=0} = i \frac{2 \pi}{\bar{\lambda}} \bFz(\brc). 
\end{equation} 
\noindent (This same equation, obtained differently, can be found in Ref. \cite{barbastasis}).

We finally obtain the pair of equations:
\begin{equation}
I_z(\brc)=\frac{1}{z^2}\int{I_0(\broc) \chi(\frac{|\brc-\broc|}{z}) d^2 \broc},
\end{equation} 
\begin{equation}
\bbtz(\brc)=\frac{1}{z^3 I_z(\brc)}\int{(\brc-\broc) I_0(\broc) \chi(\frac{|\brc-\broc|}{z}) d^2 \broc}.
\end{equation}
\noindent where $\chi(\psi)=1/(1+\psi^2)$  (Eq. 5 is obtained from Eq. 2 by setting $\brd$ to $0$; Eq. 6 is obtained from Eqs. 4 and 2). These equations may be understood from a simple ray-optics interpretation, where each point at the focal plane $0$ independently emits rays whose angular distributions, upon propagation to the aberration plane $z$, become weighted by $\chi(\psi)$. We note that these equations are simple convolutions, meaning they can be computed numerically in an efficient manner. As an aside, they can be shown also to satisfy the so-called transport of intensity equation (TIE) \cite{teague}, given by $\partial_z I_z(\brc)= -\bnabla_{\brc} \cdot ( I_z(\brc) \bbtz(\brc) )$, in agreement with the generalization of the TIE to partially coherent illumination \cite{nugent,zysk,barbastasis}.

Equations 5 and 6 are one of the main results of this paper. They provide an estimate of the wavefront at plane $z$ in the absence of the phase screen, based only on a measurement of the arbitrary object distribution $I_0(\bro)$ provided by the science camera (and a knowledge of $z$). With the additional measurement of $\bbt(\br)$ provided by our PAW sensor, we are now equipped to estimate $\bbta(\br)=\bbt(\br)-\bbtz(\br)$, corresponding to the aberrations introduced by the phase screen itself. Once estimated, $\bbta(\br)$ can be directly compensated by AO. In the case of conjugate AO, this involves simply applying the opposite (or phase-conjugate) aberrations to the correction element \cite{conjugate}, as we demonstrate experimentally below.

\section{Experimental method}

Our experimental setup is illustrated in Fig. \ref{fig:setup}. Though our setup is generalizable to fluorescence imaging, we consider here only trans-illumination imaging for simplicity.  A red LED (660nm, Thorlabs) followed by a condenser lens (Olympus) provide K\"{o}hler trans-illumination, here partially coherent (more on this later). Imaging to the science camera  (Thorlabs DCC1545M CMOS, pixel size \unit{5.2}{\micro\meter}) is provided by three 4f relays in series, where the imaging optical path is displayed in red. The total imaging magnification is $4.6\times$, with numerical aperture $0.46$ NA defined by the pupil of the $20\times$ objective (Olympus UMPlanFL).

To introduce aberrations in the imaging path, we inserted a phase screen a distance of $z\approx \unit{500}{\micro\meter}$ from the focal plane. This phase screen consisted of a photoresist film on a microscope coverslip patterned into a 2D sinusoidal array of peak-to-valley height \unit{3.5}{\micro\meter}, and period \unit{300}{\micro\meter}, as verified independently by a white-light interferometer (Zygo NT6000). As shown below, these aberrations were sufficient to significantly degrade the imaging quality of our microscope.

To compensate for these aberrations, we use the strategy of conjugate AO. A deformable mirror (Boston Micromachines Corp. MultiDM, 140 actuators in a square $12\times12$ array without the corner actuators, \unit{400}{\micro\meter} actuator pitch) is inserted in a plane conjugate to the phase screen, tilted off-axis somewhat to enable a separation of the reflected light. The operation of our conjugate AO setup is similar to that described in Ref. \cite{conjugate} except that instead of using iterative image-based AO to determine the wavefront correction (along with the requirement this imposes of guide stars in the sample), here we use sensor-based AO to directly measure the wavefront correction (no guide stars required).   

The wavefront sensor in our case is a PAW sensor comprised of a main lens and a quatrefoil lens that projects four oblique-detection images $I_{1...4}$ onto the PAW camera (Photonfocus MV1-D2080-160-CL, pixel size \unit{8}{\micro\meter}) (see Refs. \cite{paw,roman} for details). The optical path for the wavefront sensing is displayed in green (Fig. \ref{fig:setup}). The PAW sensor here measures the wavefront at the DM plane, which, in turn, is conjugate to the phase screen plane. It thus senses the composite aberrations due to the phase screen and DM combined, as characterized by the local tilt angles $\Theta(\br)_{x,y}=-\psi_c (I_1 \pm I_2 -I_3 \mp I_4)/\Sigma I_i$, where $\psi_c$ corresponds to the  soft cutoff in the angular range of illumination angles as defined by $\chi(\psi)$ in Eqs. 5 and 6 (see Discussion). Ideally, the aberrations induced by the phase screen and DM should cancel one another, and non-aberrrated imaging at the science camera should be restored, which is the goal of conjugate AO. When this happens, the residual aberrations measured by the PAW sensor should be those characterized by the wavefront tilts $\bbtz(\br)$ alone. 

In practice, the actual approach to this ideal is performed by closed-loop feedback. The PAW sensor provides a measure of the tilt angles $\bbt(\br)=\bbtz(\br)+\bbta(\br)$, where $\bbta(\br)$ arises here from the composite phase-screen/DM aberrations. An estimate of $\bbtz(\br)$ is obtained from Eqs. 5 and 6, based on the image of $I_0(\bro)$ provided by the science camera. This image is initially blurred because of the presence of the phase screen, and thus the estimate of $\bbtz(\br)$ cannot be expected to be accurate on the first try. Nevertheless, by subtracting $\bbtz(\br)$ from $\bbt(\br)$, we obtain an initial measure of $\bbta(\br)$, which is then driven toward zero by DM control (see below). The procedure is repeated in a closed loop manner, where the estimate of $\bbtz(\br)$ becomes successively improved upon each iteration as the science-camera image becomes progressively de-blurred. As we show below, loop convergence occurs rapidly after only a few iterations.        

A remaining detail is the method we use for DM control. This is a standard method generally used with SH wavefront sensing, which requires a pre-calibration of the DM to characterize the link between the control signals $\dV$ applied to the DM actuators and the resultant tilt angle map $\btp$ produced at the PAW sensor \cite{tyson}. This link is written as $\btp=\bf{M} \dV$, where $\dV$ is a vector of size equal to the number of DM actuators (here 140) and $\btp$ is a vector of size equal to twice the number of calculated pixels in the PAW wavefront reconstruction (twice because of components in $x$ and $y$). Once the calibration matrix $\bf{M}$ has been established actuator by actuator (done prior to imaging), AO can be performed. The actual control signals applied to the DM during closed-loop feedback, intended to drive $\bbta(\br)$ toward zero, are given by $\dV^{(n+1)}=\dV^{(n)}-g \bMp \bbta$, where $\bMp$ is the pseudo-inverse of $\bf{M}$, $g$ is a feedback gain (of order unity), and $n$ is a feedback iteration number. 

\section{Results}

\begin{figure}[h]
	\centering
	\makebox{\includegraphics[width=0.44\linewidth]{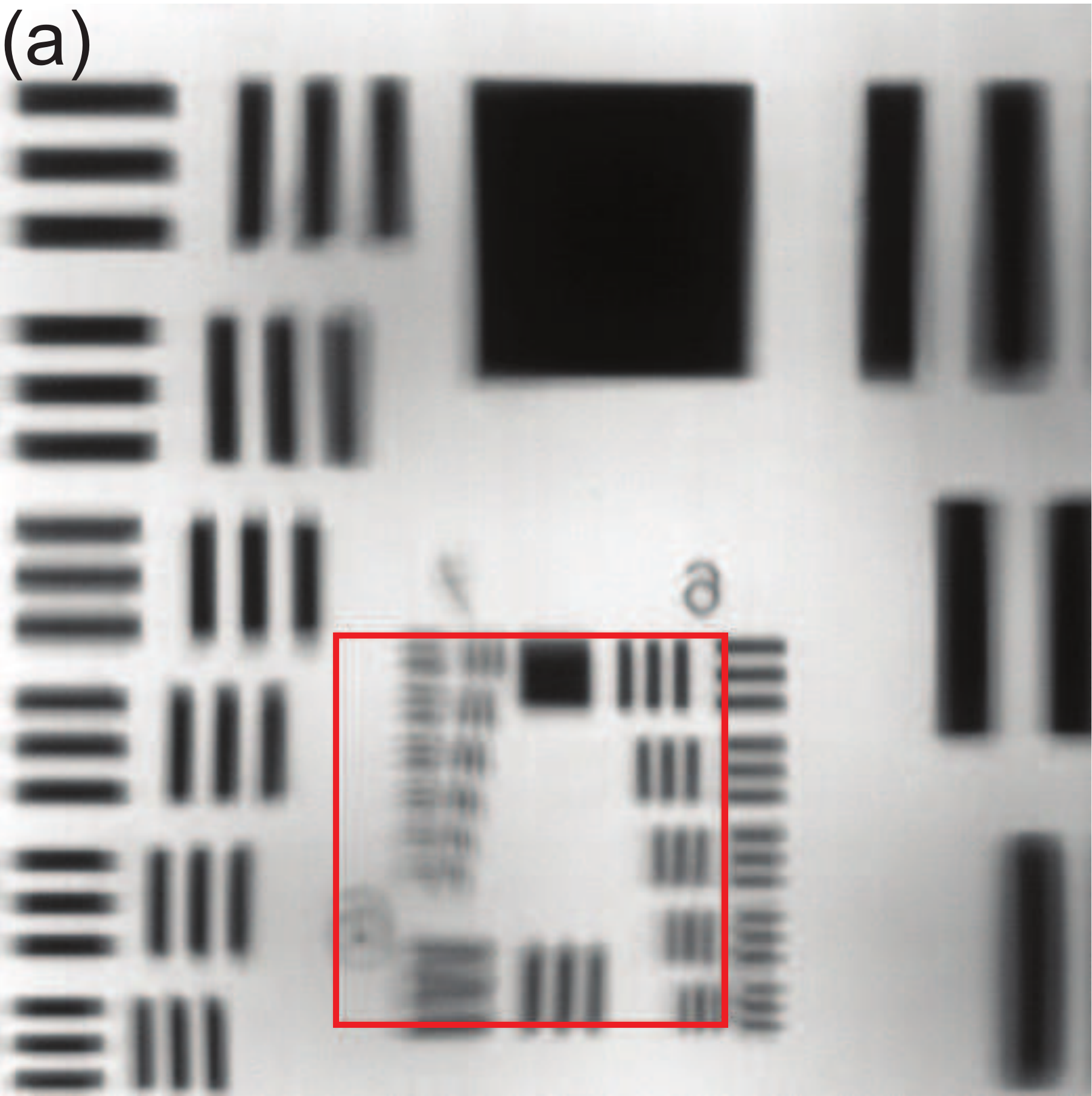}}	
	\makebox{\includegraphics[width=0.44\linewidth]{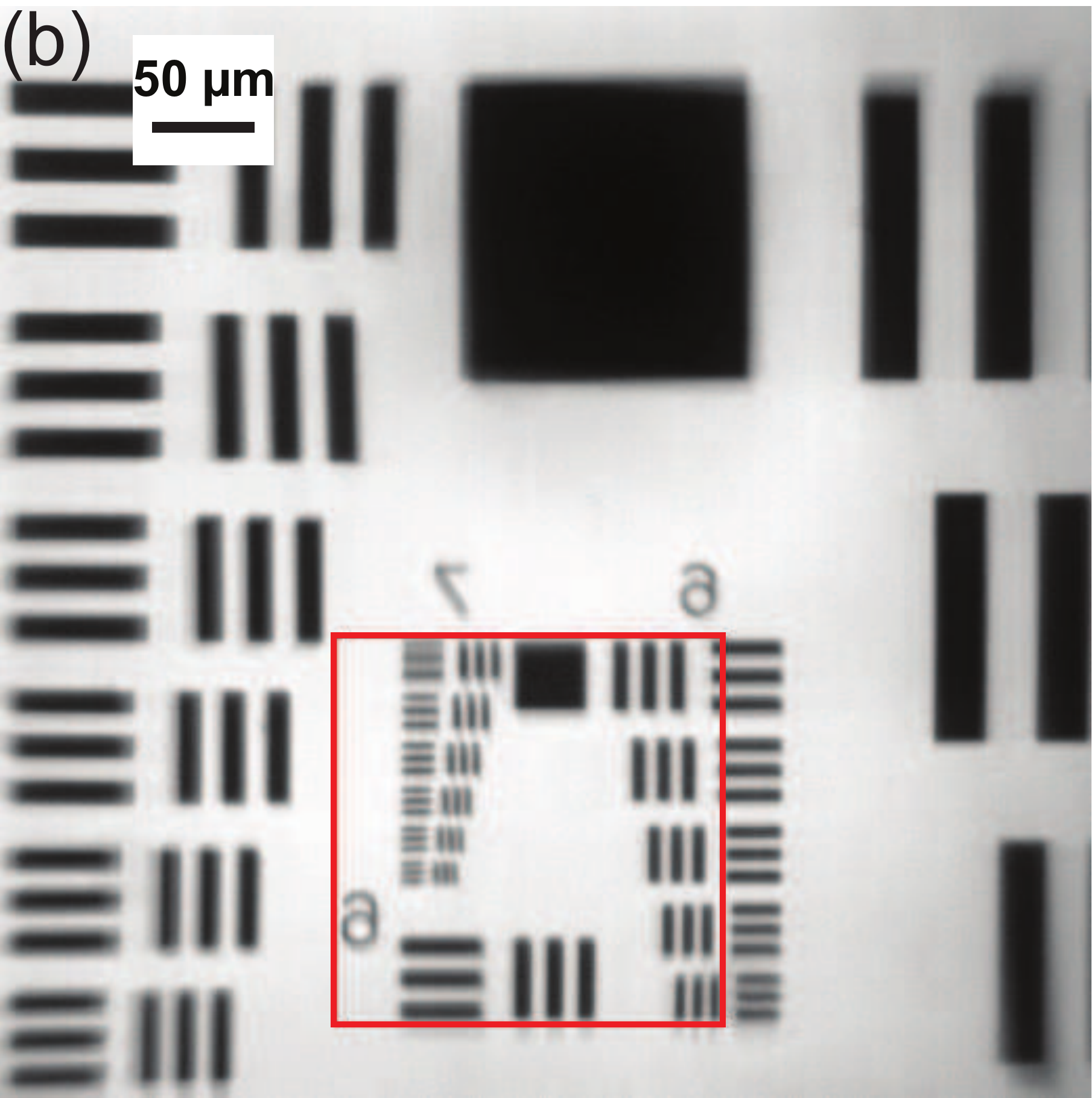}}
	\makebox{\includegraphics[width=0.44\linewidth]{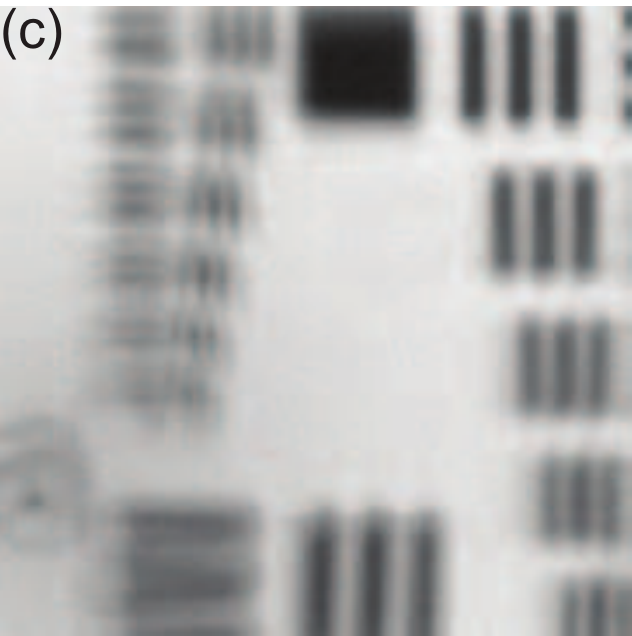}}
	\makebox{\includegraphics[width=0.44\linewidth]{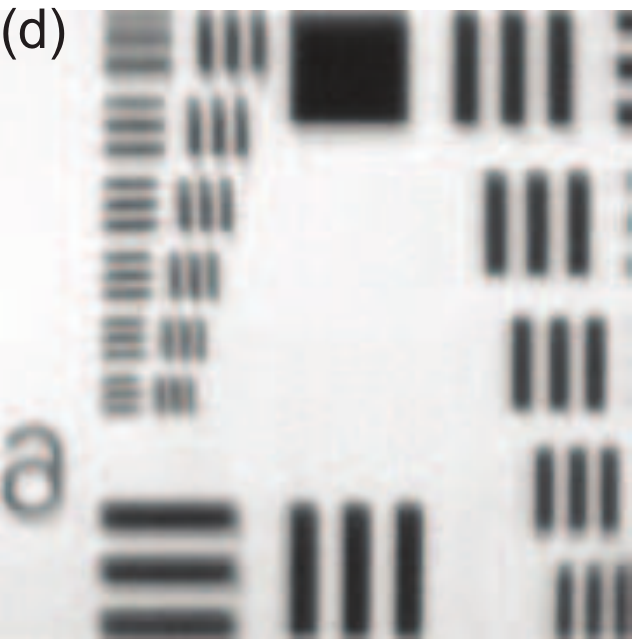}}
	\caption{Aberrated images of a 1951 USAF target sample without (a) and with (b) AO correction, and corresponding highlighted zooms (c,d).}
	\label{fig:USAF}
\end{figure}

To begin, we used a 1951 USAF calibration target as a sample. An aberrated image on this target is shown in Fig. \ref{fig:USAF}a. This image was degraded by the phase screen, albeit unevenly. For example, the smaller features of the target (zoomed inset) are particularly degraded and largely indistinguishable. Of note is the fact that the sample here is extended across the entire imaging FOV. Moreover, it is non-symmetric and highly non-uniform, presenting large intensity swings spanning close to the full dynamic range of the science camera. Despite these extended, large, non-uniform intensity swings, our sensor-based method of conjugate AO was able to substantively improve imaging quality using the illumination from the sample alone, without any additional requirement of localized guide stars, etc. The improvement was attained rapidly, in only a few feedback iterations. 

Also evident is one of the key advantages of conjugate AO over standard (pupil) AO, namely that the correction FOV is large, here spanning almost the entire surface area of the DM projected onto the sample (discounting the peripheral actuators, the active surface area is $10\times10$ actuators, corresponding to \unit{540\times540}{\micro\meter\squared} at the sample).

\begin{figure}[h]
	\centering
	\makebox{\includegraphics[width=0.44\linewidth]{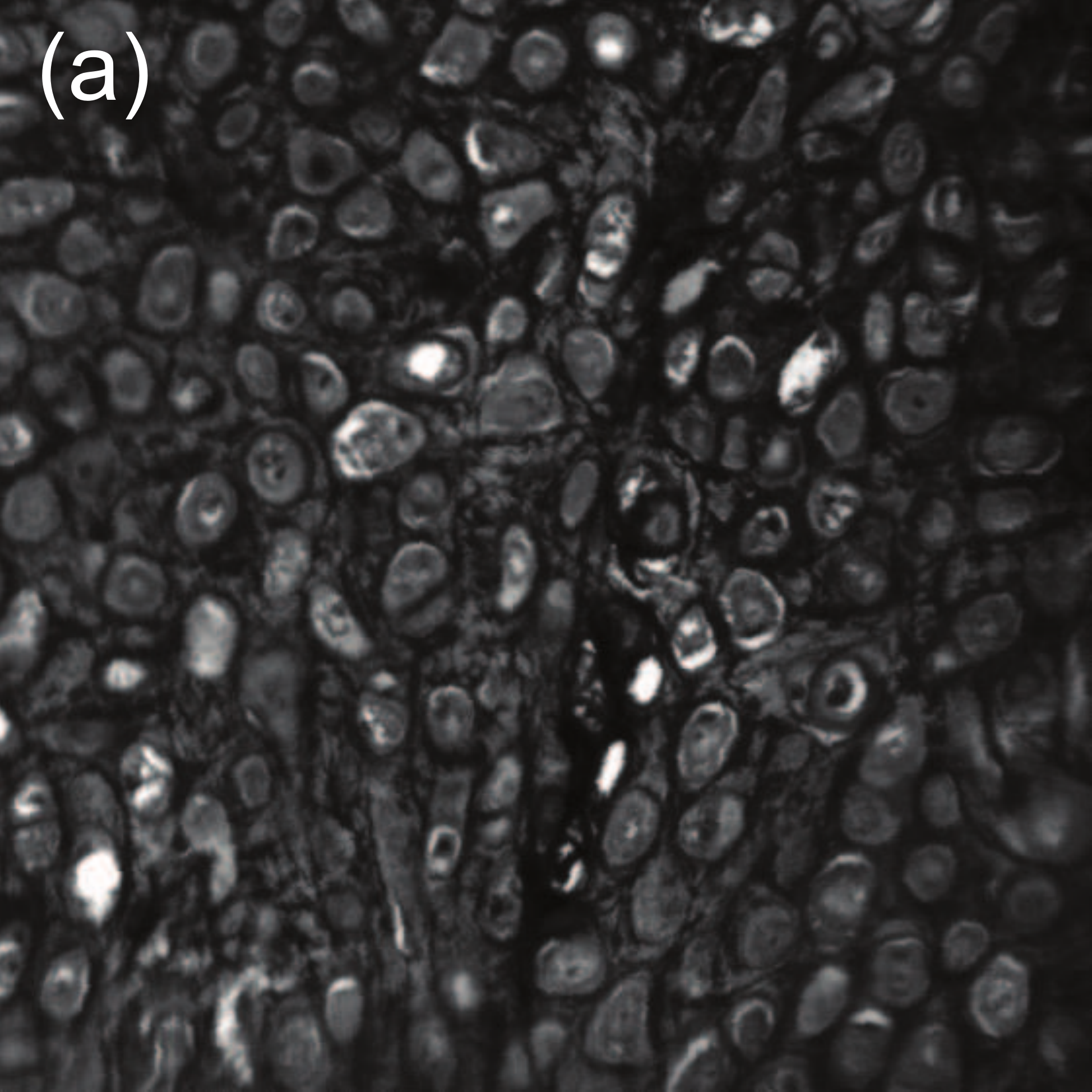}}	
	\makebox{\includegraphics[width=0.44\linewidth]{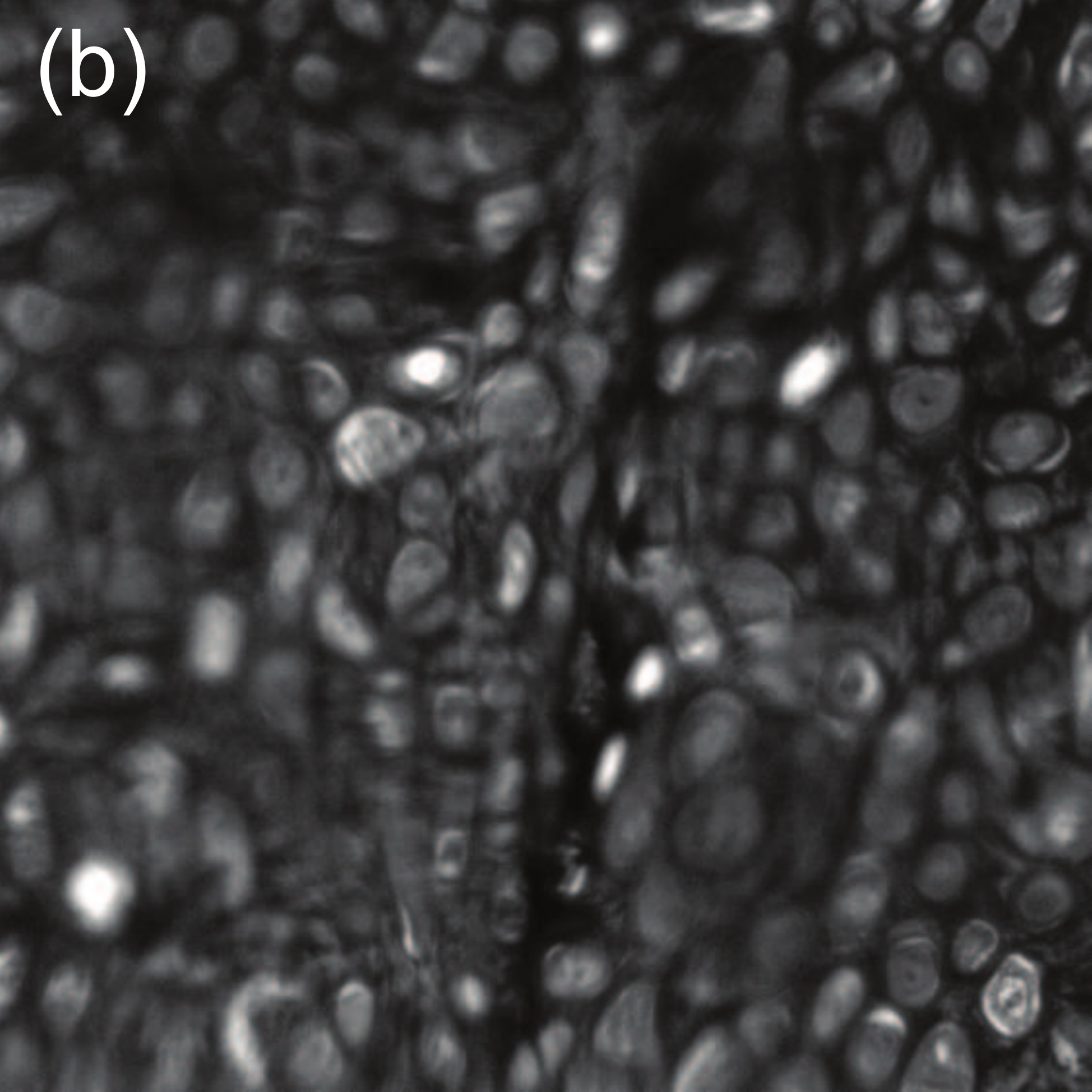}}
	\makebox{\includegraphics[width=0.44\linewidth]{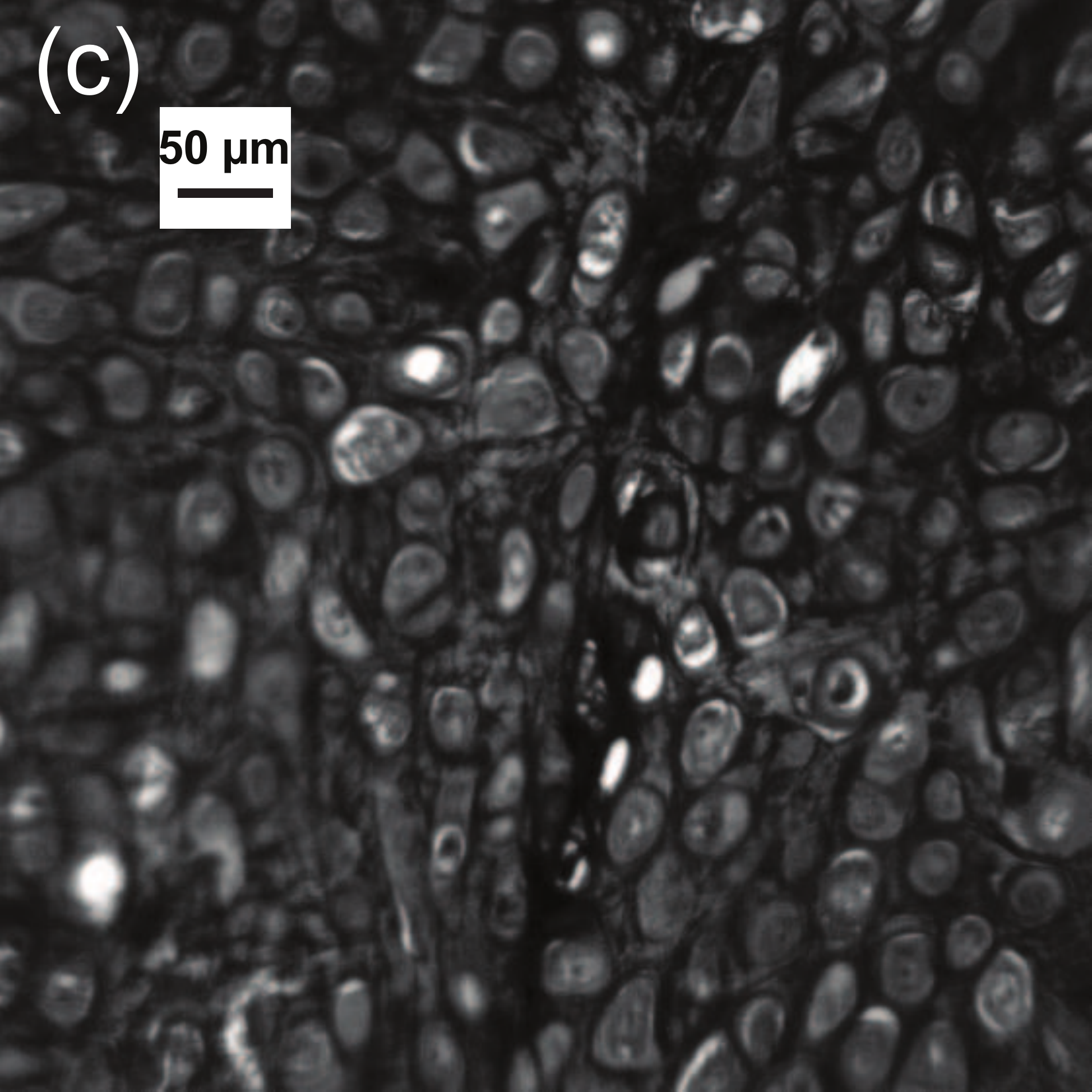}}
	\makebox{\includegraphics[width=0.44\linewidth]{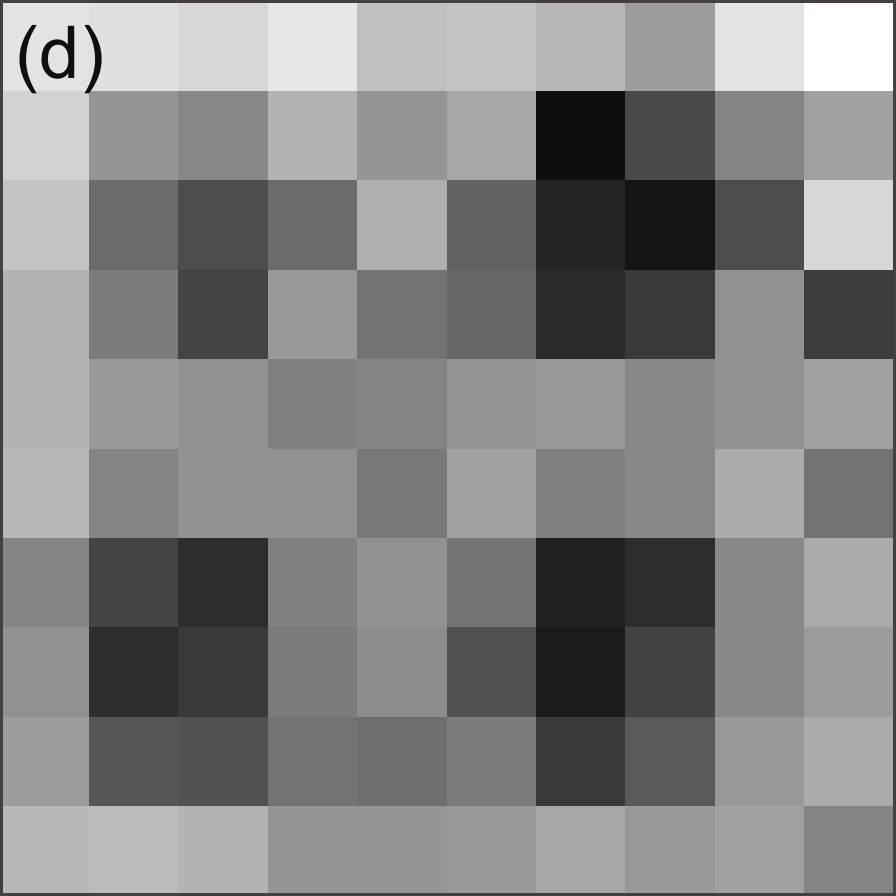}}
	\captionsetup{singlelinecheck=off}
	\caption{Images of mammal elastic cartilage with no aberrations (a: no phase screen, DM flat), uncorrected (b: phase screen, DM flat), and corrected (c: phase screen, AO on). Optimized DM actuator pattern is shown in panel (d). Note apparent periodic structure corresponding to the negative of the phase screen structure.}
	\label{fig:tissue2}
\end{figure}

As a second demonstration, we used a Verhoeff's stained mammal elastic cartilage as a sample (Carolina Biological Supply Co.). Again, sensor-based conjugate AO was able to improve image quality, almost to the level of a reference image acquired in the absence of the phase screen and with the DM replaced by a flat mirror (though some errors occur near the DM periphery, see Fig. \ref{fig:tissue2}) . Also shown is the final wavefront correction pattern applied to the DM. As expected, this has converged to roughly the negative of the 2D sinusoidal array wavefront aberrations presented by the phase screen.

\begin{figure}[h]
	\centering
	\makebox{\includegraphics[width=0.9\linewidth]{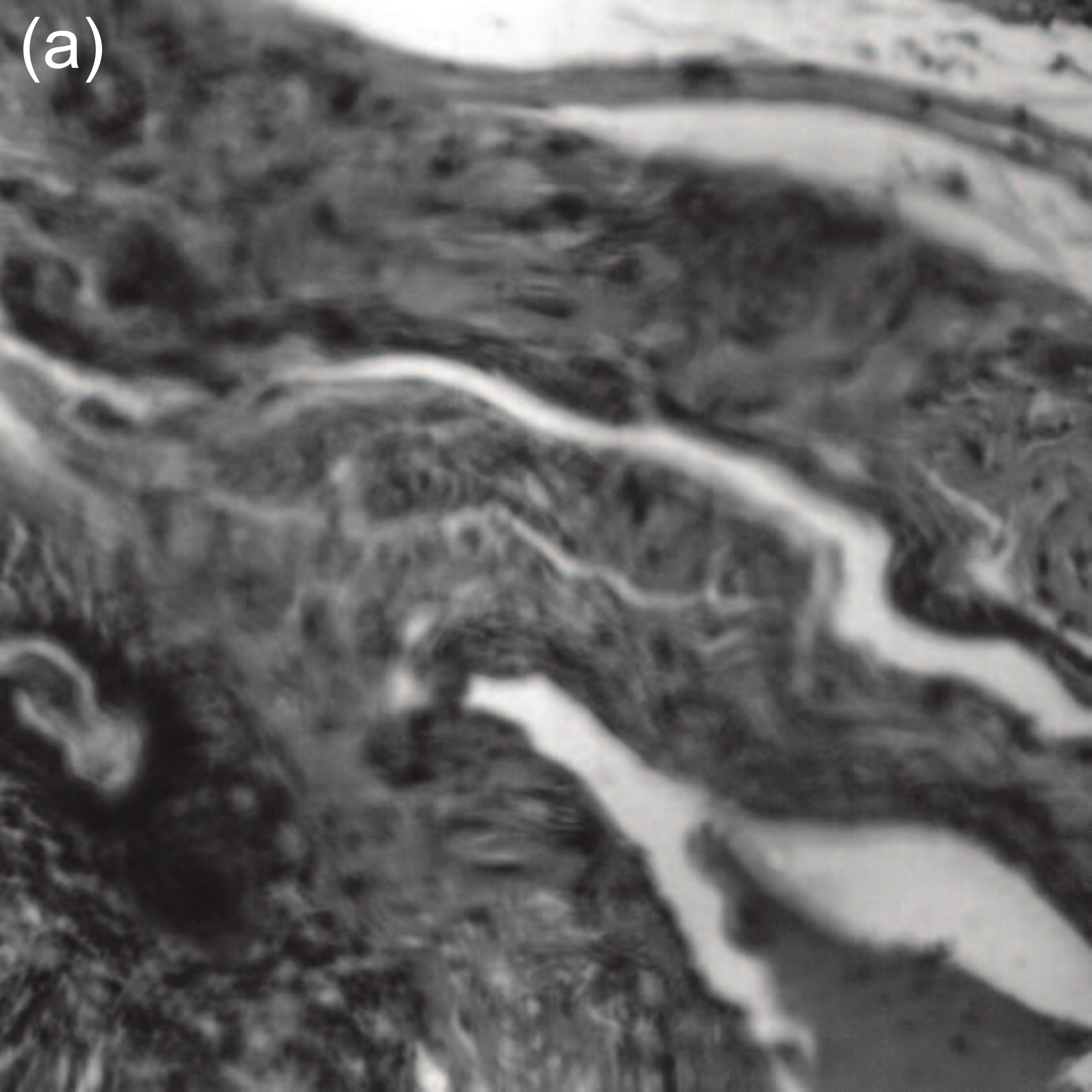}}	
	\makebox{\includegraphics[width=0.9\linewidth]{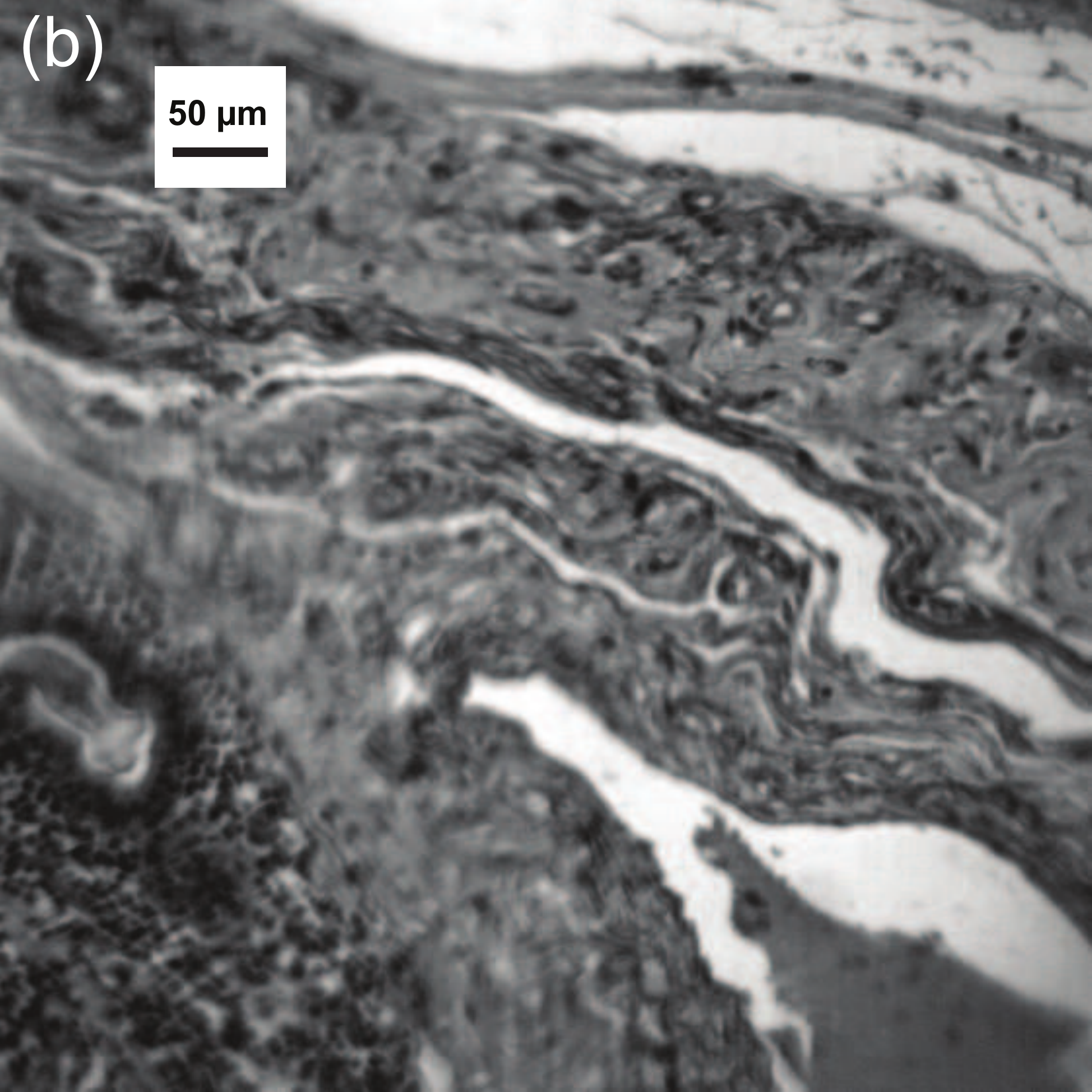}}
	\captionsetup{singlelinecheck=off}
	\caption{Aberrated images of mammal elastic cartilage without (a) and with (b) AO correction. See also Media 1 showing a video of (b) as the sample and aberrations are sporadically translated.}
	\label{fig:tissue1}
\end{figure}

For the final demonstration, we imaged another region of the elastic cartilage sample (Carolina Biological Supply Co.), without (Fig. \ref{fig:tissue1}a) and with (Fig. \ref{fig:tissue1}b) sensor-based conjugate AO correction. Once again, image quality is improved. A metric that can be used to characterize image improvement is the normalized rms error, defined by $\sqrt{\langle(I_{ao}-I_o)^2\rangle}/\langle I_o \rangle$, where  the brackets denote an average over image pixels, and $I_{ao}$ and $I_o$ are respectively the image obtained with AO correction and the reference image obtained in the absence of aberrations (co-registered). A plot of this rms error is shown as a function of feedback iteration number for various values of the feedback gain $g$ (Fig. \ref{fig:convergence}). When $g$ is too small, the convergence rate is modest; when $g$ is too large, the system is driven into oscillation. An optimal feedback gain that leads to the fastest rate of stable convergence is found to be close to $1$. At this gain setting about 4 or 5 iterations suffice to achieve near-maximal AO correction. In our case, the time required per iteration was roughly 1s, limited by the speed of our Matlab software (hardware limitations such as the 34 fps of our PAW sensor camera and the 30kHz update rate of our DM were not a bottleneck). It should be noted that we made no special efforts to optimize the speed of our software, which we expect could be significantly increased by proper streamlining or operation with a graphical processing unit (GPU).

\section{Discussion}

In summary, we have demonstrated the feasibility of sensor-based AO in a widefield microscope configuration (as opposed to the much more common scanning microscope configuration). Our technique makes use of a wavefront sensor that, with the help of the science camera, requires no guide stars and uses the arbitrarily distributed sample itself as the illumination source. A key advantage of sensor-based over image-based AO is that it provides a direct measure of wavefront rather than a measure obtained through iterative trial and error which, in the absence of guide stars, often fails to converge correctly. Another key advantage is that it has the potential to be much faster, by orders of magnitude.

\begin{figure}[ht]
	\centering
	\makebox{\includegraphics[width=0.49\linewidth]{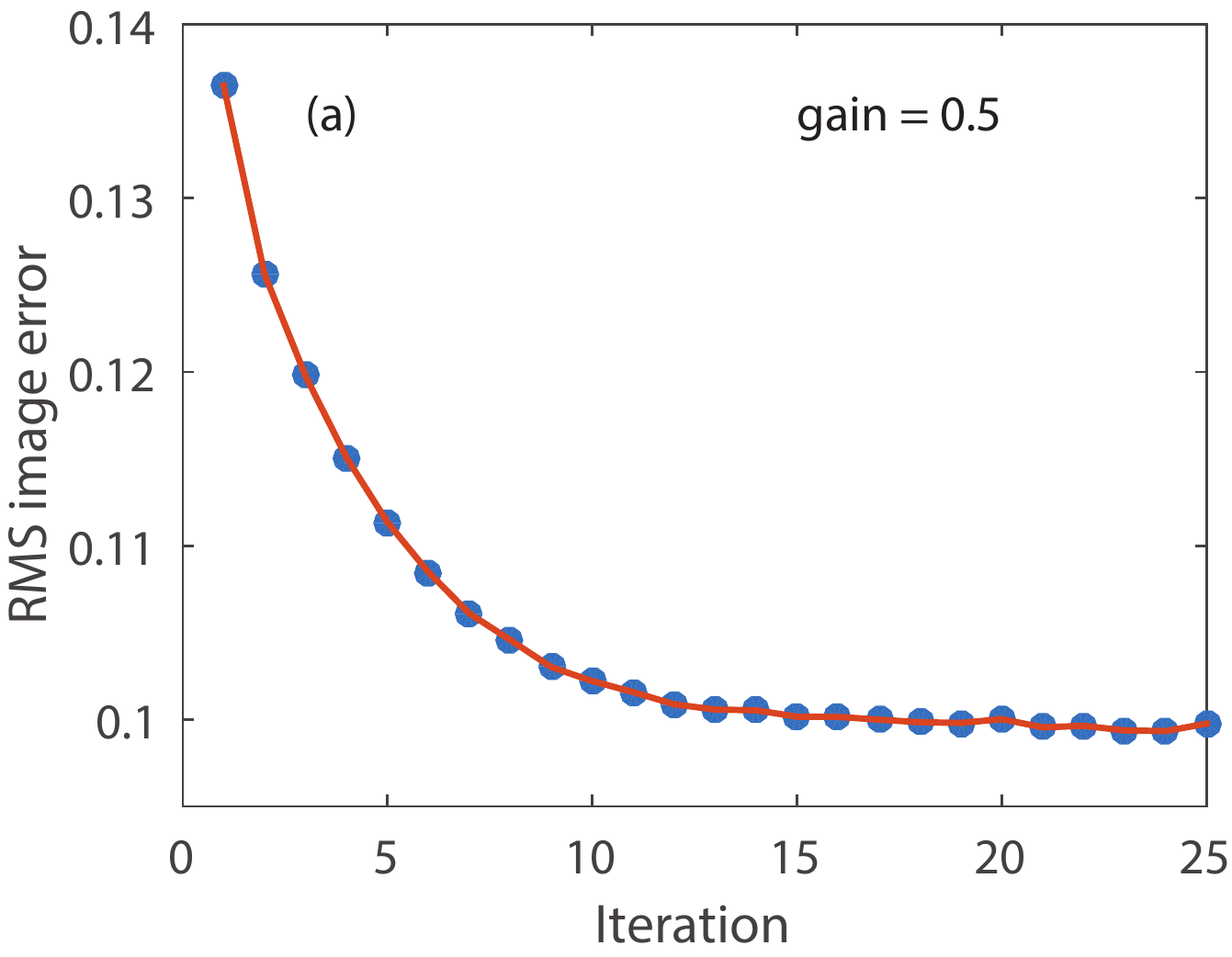}}	
	\makebox{\includegraphics[width=0.49\linewidth]{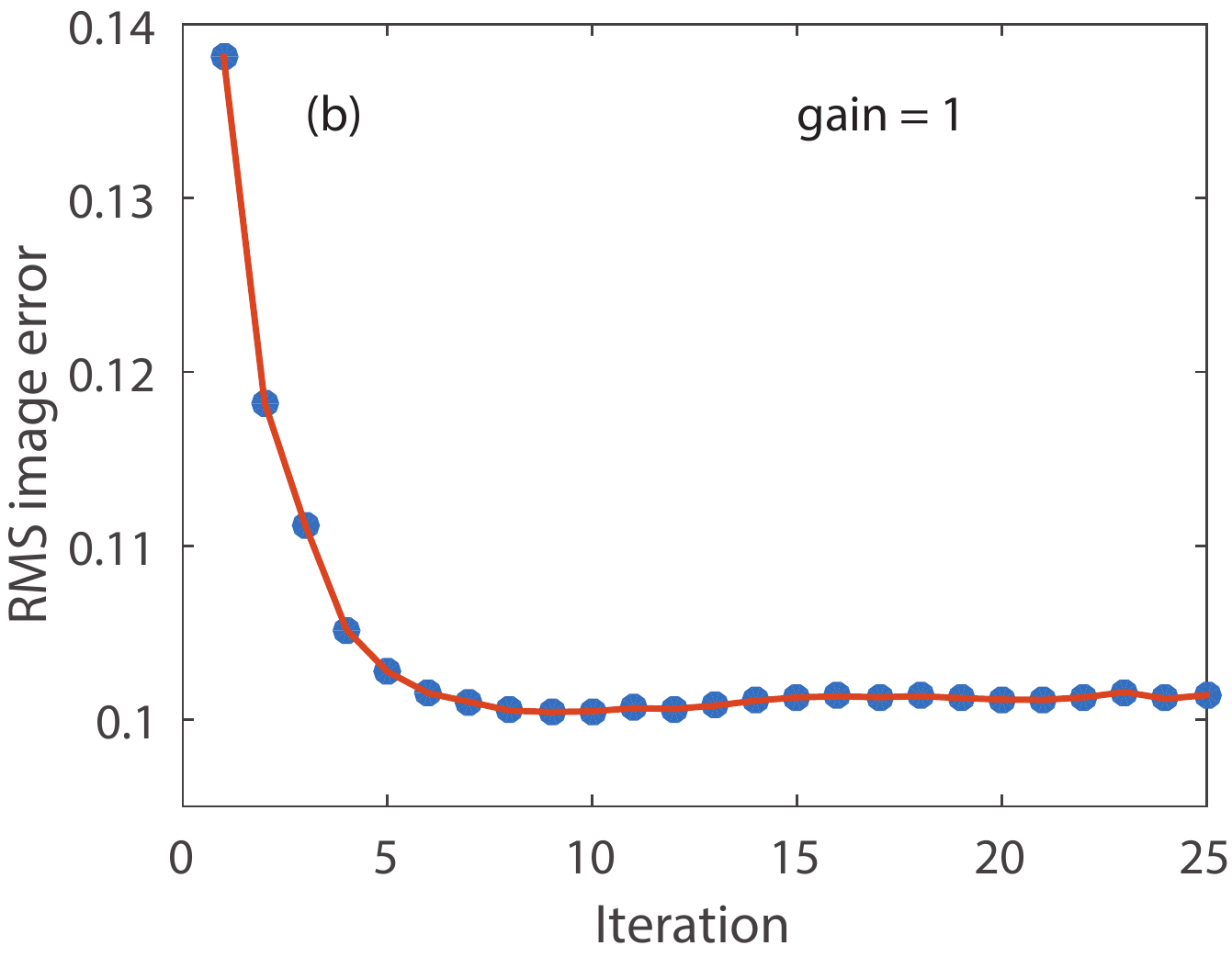}}
	\makebox{\includegraphics[width=0.49\linewidth]{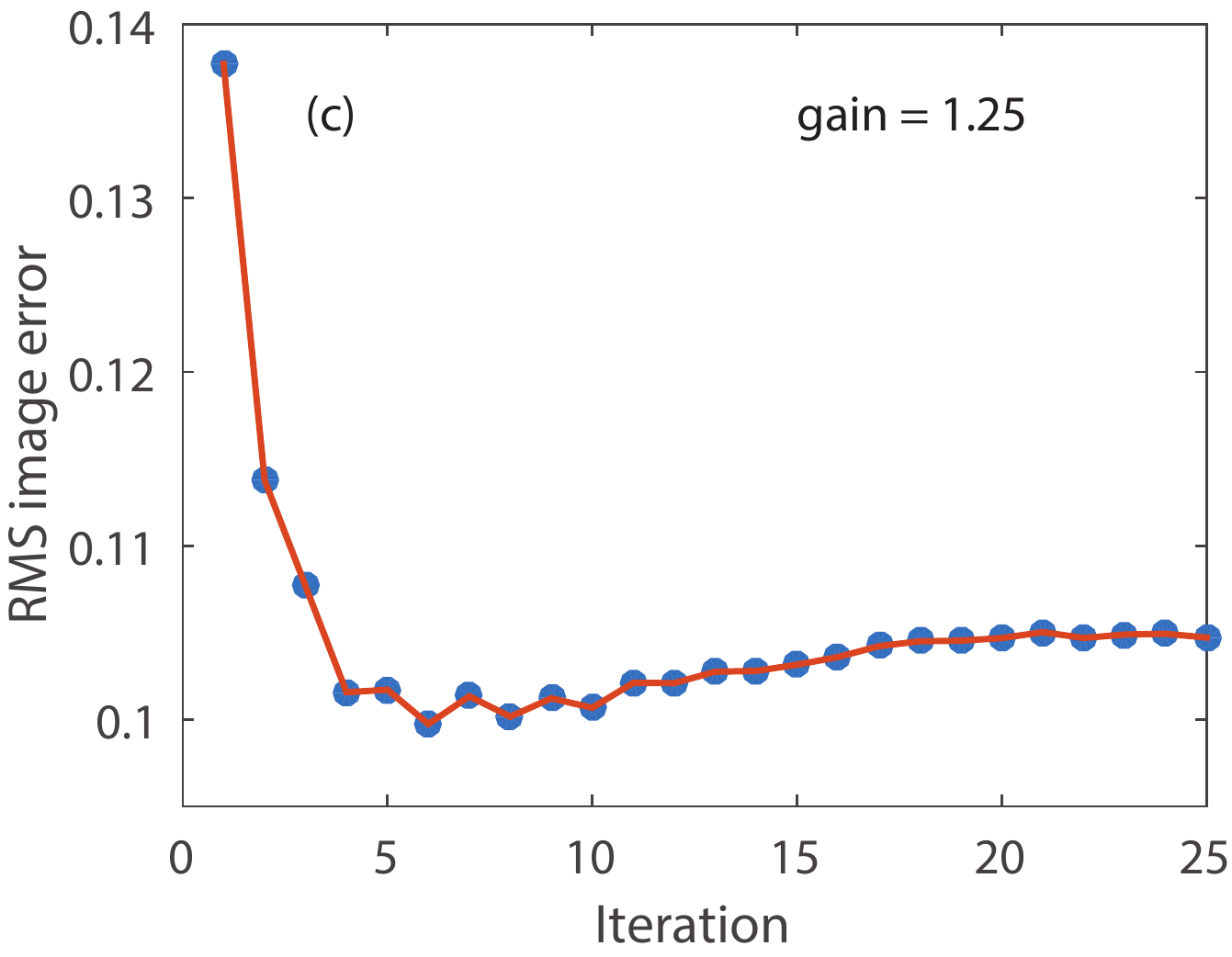}}
	\makebox{\includegraphics[width=0.49\linewidth]{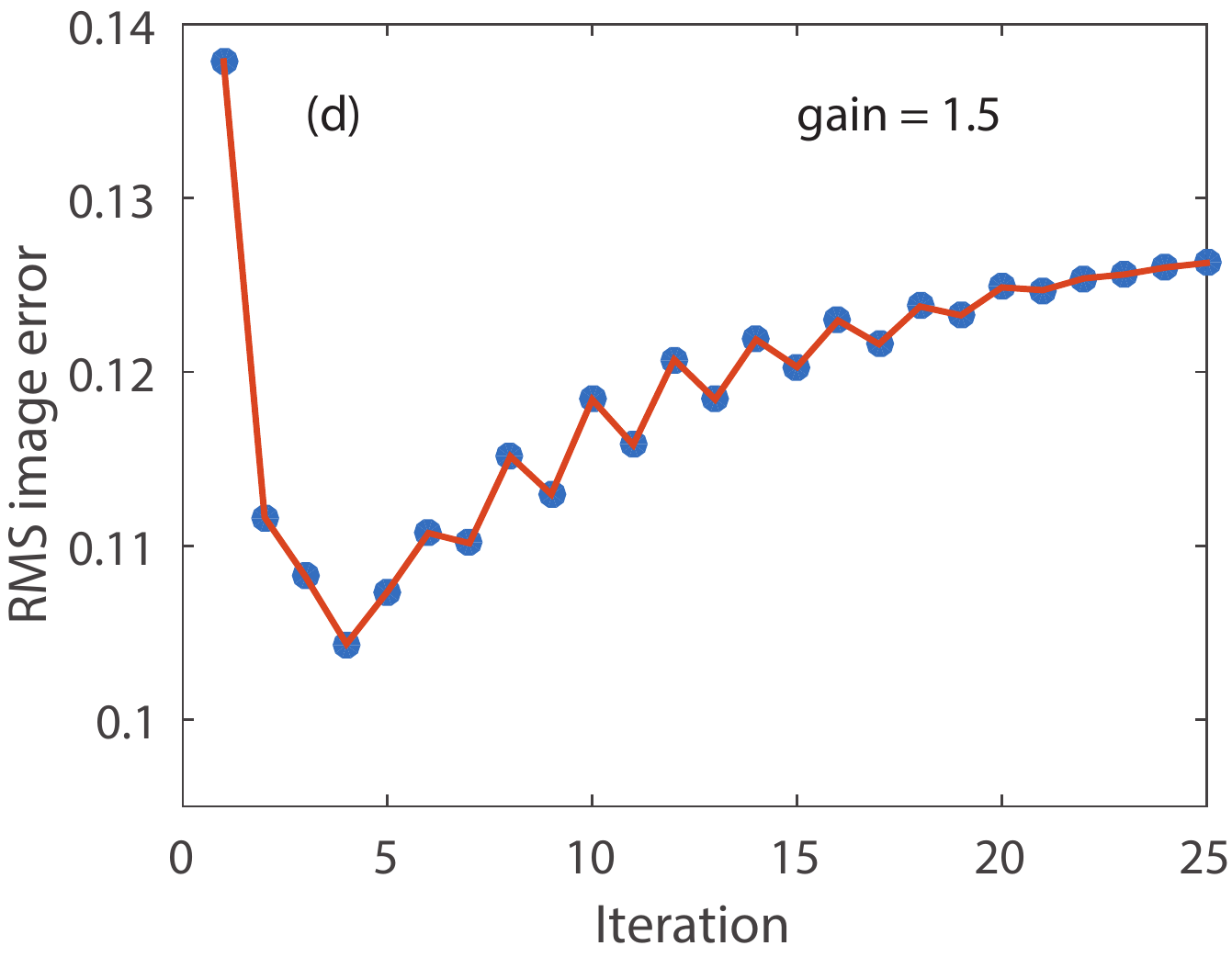}}
	\captionsetup{singlelinecheck=off}
	\caption{Convergence of AO correction as a function of feedback iteration. Normalized rms image error is shown for different feedback gains $g$. }
	\label{fig:convergence}
\end{figure}

Despite these advantages, some words of caution are in order. A first caution comes from limitations in the AO correction element. To properly cancel wavefront distortions produced at the aberration plane, the conjugate correction element must provide commensurate spatial resolution and dynamic range. Here our correction element was a DM of only modest resolution (number of actuators) and dynamic range (stroke), meaning our system was able to operate only with aberrations that were relatively long range and weak. Though we designed our aberrations photo-lithographically to be within the range of our DM specifications, limitations in these specifications may still have been responsible for residual errors, as manifested in Fig. \ref{fig:convergence}. (Another cause of residual errors might be the tilt angle of our DM with respect to the aberration plane, undermining proper conjugation.)

A second caution comes from a limitation of PAW sensing itself. While PAW provides a significantly larger dynamic range than SH wavefront sensing, its dynamic range still remains bounded. Specifically, PAW operates under the condition that the NA used to illuminate the wavefront plane of interest (here the aberration plane) be smaller than the NA used to detect this plane \cite{paw}. In cases of sample trans-illumination where the illumination NA can be readily controlled by an aperture stop, this condition can be easily met. In these cases, the angular distribution function $\chi(\psi)$ in Eqs. 5 and 6, which is applicable to illumination derived from a spatially incoherent source a distance $z$ from the plane of interest, should be replaced by a narrower distribution function applicable to a partially coherent source. For example, in our demonstration experiments, our detection NA was 0.46 and we adjusted our illumination NA to be about 0.2. Such an adjustment of illumination NA would be more difficult in the case of fluorescence imaging. In such a case, the angular distribution of $\psi$ at the aberration plane becomes limited either by the range of $|\brc-\broc|/z$, as determined by the distribution and distance of the fluorescent sources, or by the range of $\chi(\psi)$, which imparts a soft cutoff to $\psi$ even in conditions where $|\brc-\broc|/z$ is large. This cutoff occurs at an NA of about 0.7 (in air), meaning that the detection NA should be higher than this value in the extreme case of very extended fluorescent sources not far from the aberration plane. However, in the event that such high detection NA is impractical one must resort instead to controlling the range of $|\brc-\broc|/z$, for example by limiting the spatial extent of the fluorescent sources with a field stop in the excitation optics.

A third caution comes from the assumptions made throughout this work. Specifically, we only considered a very simplified geometry where the sample and aberration planes are planar and separated by a well defined distance $z$. Such a geometry may be encountered in practice, for example when imaging a fluorescent layer situated behind an aberrating interface (e.g. in light sheet microscopy, retinal imaging, etc.). In general, however, both the sample and aberrations may be axially distributed. Our technique, therefore, should be generalized to accommodate both out of focus sources and multiple aberration planes. For example, it is not clear to what degree the FOV benefits of conjugate AO are preserved in the case of multiple aberration planes. Certainly multi-conjugate AO can help preserve these benefits, as is well known from astronomical imaging \cite{beckers}. Numerical simulations \cite{sedat,cui_wf} have also suggested that benefits subsist even in the case of conjugate AO with a single correction element. Such benefits, however, remain to be demonstrated experimentally in microscopy applications with thick samples.

As such, the work presented in this paper should be considered as preliminary only. Nevertheless, the field of AO applied to microscopy is advancing rapidly. Given the potential benefits of widefield, sensor-based conjugate AO, we hope the general strategy presented here will constitute a step forward in this advance.   

\section*{Funding information}
Support for this project was provided in part by the National Science Foundation Industry/University Cooperative Research Center for Biophotonic Sensors and Systems, the Boston University Center for Systems Neuroscience, and by the National Institute of Health. 

\section*{Acknowledgments}
T. Bifano acknowledges a financial interest in Boston Micromachines Corporation.\\

\noindent See Supplement 1 for supporting content.

%
%
%
%



\begin{thebibliography}{1}

\bibitem{tyson}
R. Tyson, \textit{Principles of Adaptive Optics, Third Edition}, Series in Optics and Optoelectronics, CRC Press (2010).  


\bibitem{kubby_book}
J. A. Kubby, Ed., \textit{Adaptive Optics for Biological Imaging}, CRC Press (2013). 

\bibitem{booth_review}
M. J. Booth, ``Adaptive optical microscopy: the ongoing quest for a perfect image." Light: Sci. and Appl. 3, e165 (2014).

\bibitem{beckers}
J.M. Beckers, ``Increasing the size of the isoplanatic patch within multiconjugate adaptive optics,"  Proceedings of
European Southern Observatory Conference and Workshop on Very Large Telescopes and Their Instrumentation, ESO Conference and
Workshop Proceedings 30,  693-703 (1988).


\bibitem{sedat}
Z. Kam, P. Kner, D. Agard, and J. W. Sedat, ``Modelling the application of adaptive optics to wide-field microscope live imaging", J. Microsc. 226, 33-42 (2007).

\bibitem{booth}
R. D. Simmonds and M. J. Booth, ``Modelling of multi-conjugate adaptive optics for spatially variant aberrations in microscopy," J. Opt. 15, 094010 (2013).

\bibitem{cui_wf}
T.-W. Wu and M. Cui, ``Numerical study of multi-conjugate large area wavefront correction for deep tissue microscopy," Opt. Express 23, 7463-7470 (2015). 

\bibitem{retina}
J. Thaung, P. Knutsson, Z. Popovic, and M. Owner-Petersen, ``Dual-conjugate adaptive optics for wide-field high-resolution retinal imaging," Opt. Express 17, 4454-4467 (2009).

\bibitem{conjugate}
J.~Mertz, H.~Paudel, T.~G. Bifano, ``Field of view advantage of conjugate adaptive optics in microscopy applications," Appl. Opt. 54, 3468-3506 (2015).

\bibitem{girkin}
P. N. Marsh, D. Burns, and J. M. Girkin, ``Practical implementation of adaptive optics in multiphoton microscopy." Opt. Express 11, 1123-1130 (2003).

\bibitem{debarre}
D. Débarre, E. J. Botcherby, T. Watanabe, S. Srinivas,
M. J. Booth, and T. Wilson, ``Image-based adaptive optics for two-photon microscopy", Opt. Lett. 34, 2495-2497 (2009).

\bibitem{ji}
C. Wang, R. Liu, D. E. Milkie, W. Sun, Z. Tan, A. Kerlin,
T.-W. Chen, D. S. Kim, and N. Ji, ``Multiplexed aberration measurement for deep tissue imaging in vivo." Nat. Meth. 11, 1037-1040 (2014). 

\bibitem{mosk}
I. M. Vellekoop and A. P. Mosk, ``Focusing coherent light through opaque strongly scattering media." Opt. Lett. 32, 2309-2311 (2007).

\bibitem{cui}
J. Tang, R. N. Germain, and M. Cui, ``Superpenetration optical microscopy by iterative multiphoton adaptive compensation technique." Proc. Nat. Acad. Sci. U.S.A. 109, 8434-8439 (2012).  

\bibitem{miller}
 J. Liang, D. R. Williams, and D. T. Miller, ``Supernormal vision and high-resolution retinal imaging through adaptive optics." J. Opt. Soc. Am. A 14, 2884-2892 (1997).


\bibitem{wilson}
M. J. Booth, M. A. A. Neil, R. Ju\u{s}kaitis, R. and T. Wilson, ``Adaptive
aberration correction in a confocal microscope.'" Proc. Natl. Acad. Sci.
U.S.A. 99, 5788-5792 (2002).

\bibitem{denk}
M. Rueckel, J. A. Mack-Bucher, and W. Denk, ``Adaptive wavefront correction in two-photon microscopy using coherence-gated wavefront sensing," Proc. Nat. Acad. Sci. U.S.A. 103, 17137-17142 (2006).


\bibitem{kubby}
X. Tao, J. Crest, S. Kotadia, O. Azucena, D. C. Chen, W. Sullivan, and J. Kubby, ``Live imaging using adaptive optics with fluorescent protein guide-stars," Opt. Express 20, 15969-15982 (2012).

\bibitem{betzig}
K. Wang, D. E. Milkie, A. Saxena,
P. Engerer, T. Misgeld, M. E. Bronner, J. Mumm, and E. Betzig, ``Rapid adaptive optical recovery of optimal resolution over large volumes." Nat. Meth. 11, 625-628 (2014).

\bibitem{shack}
B. C. Platt and R. Shack, ``History and principles of Shack-Hartmann wavefront sensing," J. Refract. Surg. 17, S573-S577 (2001).

\bibitem{paw}
A. B. Parthasarathy, K. K. Chu, T. N. Ford, and J. Mertz, “Quantitative phase imaging using a partitioned detection aperture,” Opt. Lett. 37, 4062–4064 (2012). 

\bibitem{roman}
R. Barankov and J. Mertz, ``Single-exposure surface profilometry using partitioned aperture wavefront imaging," Opt. Lett. 38, 3961-3964 (2013).

\bibitem{fienup}
R. G. Paxman, T. J. Schulz, and J. R. Fienup, ``Joint estimation of object and aberrations by using phase diversity," J. Opt. Soc. Am. A 9, 1072-1085 (1992).

\bibitem{allen}
L. J. Allen and M. P. Oxley, ``Phase retrieval from series of images obtained by defocus variation," Opt. Commun. 199, 65-75 (2001).

\bibitem{gureyev}
T. E. Gureyev, Y. I. Nesterets, D. M. Paganin, A. Pogany, S. W. Wilkins, ``Linear algorithms for phase retrieval in the Fresnel region. 2. Partially coherent illumination," Opt. Commun. 259, 569-580 (2006).

\bibitem{barbastasis}
J. C. Petruccelli, L. Tian, and G. Barbastathis, ``The transport of intensity equation for optical path length recovery using partially coherent illumination," Opt. Express 21, 14430-14441 (2013).

\bibitem{waller}
Z. Jingshan, L. Tian, J. Dauwels, and L. Waller, ``Partially coherent phase imaging with simultaneous source recovery," Biomed. Opt. Express 6, 257-265 (2014).

\bibitem{popescu}
G. Popescu, \textit{Quantitative phase imaging of cells and tissues}, McGraw-Hill Biophotonics, McGraw-Hill (2011).

\bibitem{slim}
Z. Wang, L. Millet, M. Mir, H. Ding, S. Unarunotai, J. Rogers, M. U. Gillette, and G. Popescu, ``Spatial light interference microscopy (SLIM)," Opt. Express 19, 1016-1026 (2011).

\bibitem{bernet}
S. Bernet, A. Jesacher, S. Fuerhapter, C. Maurer, and M.
Ritsch-Marte, ``Quantitative imaging of complex samples by spiral phase contrast microscopy," Opt. Express 14, 3792-3805 (2006).

\bibitem{arnison}
M. R. Arnison, K. G. Larkin, C. J. R. Sheppard, N. I.
Smith, and C. J. Cogswell, ``Linear phase imaging using differential interference contrast microscopy," J. Microsc. 214, 7-12 (2004).

\bibitem{monneret}
P. Bon, G. Maucort, B. Wattellier, and S. Monneret, ``Quadriwave lateral shearing interferometry for quantitative phase microscopy of living cells,'' Opt. Express 17, 13080-13094 (2009).

\bibitem{iglesias}
I. Iglesias, ``Pyramid phase microscopy,'' Opt. Lett. 36, 3636-3638 (2011).

\bibitem{ishimaru}
A. Ishimaru, \textit{Wave Propagation and Scattering in Random Media}, Wiley-IEEE Press (1999).

\bibitem{bw}
M. Born and E. Wolf, \textit{Principles of Optics: Electromagnetic Theory of Propagation, Interference and Diffraction of Light}, Cambridge University Press (1999). 

\bibitem{teague}
M. R. Teague, “Deterministic phase retrieval: a Green’s function solution,” J. Opt. Soc. Am. A 73, 1434–1441 (1983).

\bibitem{nugent}
D. Paganin and K. Nugent, “Noninterferometric phase imaging with partially coherent light,” Phys. Rev. Lett.
80, 2586–2589 (1998).

\bibitem{zysk}
A. M. Zysk, R. W. Schoonover, P. S. Carney, and M. A. Anastasio, “Transport of intensity and spectrum for
partially coherent fields,” Opt. Lett. 35, 2239–2241 (2010).



\end{thebibliography}
\end{document}